\documentclass[conference]{IEEEtran}
\usepackage{graphicx}
\usepackage{amssymb,amsthm, amsfonts}
\usepackage{ifthen}
\usepackage{tikz}
\usepackage{enumerate}
\usepackage{verbatim}
\usepackage{amsmath}
\usepackage{amsmath}
\usepackage{cancel}
\usepackage[normalem]{ulem}
\usepackage{float}
\usepackage{enumitem}
\usepackage{textpos}
\usepackage{url}

\newtheorem{theorem}{Theorem}

\newtheorem{remark}[theorem]{Remark}

\DeclareSymbolFont{bbold}{U}{bbold}{m}{n}
\DeclareSymbolFontAlphabet{\mathbbold}{bbold}

\newcommand{\be}{\begin{equation}}
\newcommand{\ee}{\end{equation}}
\newcommand{\ben}{\begin{equation*}}
\newcommand{\een}{\end{equation*}}
\newcommand{\ba}{\begin{align}}
\newcommand{\ea}{\end{align}}

\usetikzlibrary{arrows}
\usetikzlibrary{patterns}
\usetikzlibrary{shapes}
\usetikzlibrary{decorations}
\usetikzlibrary{decorations.footprints}
\usetikzlibrary{decorations.pathreplacing}
\usetikzlibrary{decorations.pathmorphing}

\title{Dispensing with Noise Forward in the ``Weak'' Relay-Eavesdropper Channel}
\author{
\IEEEauthorblockN{Krishnamoorthy Iyer}
\IEEEauthorblockA{Department of Electrical Engineering, Indian Institute of Technology Bombay
\\
	{\tt krishna@ee.iitb.ac.in}
}

{\color{red}}}

\begin{document}

\maketitle

\begin{abstract}

The ``weak'' relay-eavesdropper channel was first studied by Lai and El Gamal, whose achievable scheme introduced noise forwarding (NF) and used backward decoding. We suggest a novel sliding window decoding scheme with a two block decoding delay, where the relay uses compress-forward with Wyner-Ziv (WZ) binning but does not use NF. Wireless engineers will welcome the reduced decoding delay. Sliding window decoding mandates multiblock equivocation calculations;  dispensing with NF enables it. We identify nine regimes and develop a case-by-case choice of relay channel codebook and WZ bin sizes to maximize the secrecy rate. The multiblock equivocation calculations may be of independent interest.
\end{abstract}
%
%
\section{Introduction}
\label{sec:intro}
The ``weak'' relay-eavesdropper channel with a trusted relay and external eavesdropper (Eve) has been the focus of interest since 
\cite[Theorems~$3,4$]{LaiElGam08}. By ``weak'' is meant a relay that does not decode-and-forward its received sequence, but one which compresses and sends the bin index of the compressed sequence \cite{ElGamKim2012}. \cite{LaiElGam08} introduced the idea of NF and generalized NF (aka GNF), wherein the relay transmits some random codewords to confuse both the intended receiver (Bob) and Eve. Eve is affected more than Bob; leading to an increased secrecy rate. In \cite{LaiElGam08}, the relay did not perform WZ binning. A useful improvement to \cite{LaiElGam08}'s scheme was developed by 
\cite{TLSP2011}, who showed that dispensing with the requirement that \textit{given the message, Eve be able to decode the ``relay's'' codeword} leads to an increase in secrecy rate. Note that in pure NF \cite[Theorem~$3$]{LaiElGam08} and in the helping interferer \cite{TLSP2011}, the ``relay'' actually doesn't relay; it acts as an oblivious i.e., deaf helper. 

The relay channel has itself been the focus of continued research interest. 
Recently, \cite{LuoGohYanik2012} have developed a new decoding scheme for the ``weak'' relay channel. 
While no rate improvement was achieved for the canonical relay channel, they showed an improvement for multi-relay networks. In our scheme, we use the decoding technique from \cite{LuoGohYanik2012} together with ideas from \cite{TLSP2011} and \cite{WuFanXie10}. 

Both \cite{LuoGohYanik2012}, \cite{WuFanXie10} indicate that too high a compression sequence rate, 
or too low a relay channel codeword rate can reduce the message rate. \textit{This brings us to the main idea of our paper i.e., to use compression sequences themselves to interfere with and confuse Eve by choosing appropriate compression rates and relay channel codeword, equivalently, WZ binning rates}. 

Sliding window decoding uses multiblock correlations to decode, entailing that the equivocation calculation must necessarily also be multiblock. A previous paper by the author \cite{KIyer2018} also uses multiblock equivocation (MBEq) in a (dedicated and trusted) four node relay broadcast channel with two receivers, each requiring an independent message to be kept secret from the other. The relay in \cite{KIyer2018} is a ``strong'' relay that decodes-and-forwards the messages to their destinations. 

In related work, 
\cite{DaiYuMa16} have considered a ``weak'' relay-eavesdropper with long message noisy network coding (LM-NNC) scheme that also dispenses with NF as we have done by making use of the insight that judiciously choosing the number of compression sequences can impact the secrecy rate achievable. However, they do not employ Wyner-Ziv binning, and the decoding delay of LM-NNC is $B$ blocks. 
\section{Model: ``Weak'' Relay-Eavesdropper Channel}
\label{sec:rel_eve_cf}

\begin{figure}[htb]
\begin{center}
\begin{tikzpicture}[yscale=0.8, line width=1.5pt]

\coordinate (cenc) at (0.25,0);
\coordinate (cchan)at (3,0);
\coordinate (cdec1)at (6.0,0.9);
\coordinate (cdec2)at (6.0,-0.9);
\coordinate (ck1)  at (3,3);

\node (enc) at (cenc) [rectangle, draw, minimum width=1.5cm,minimum height=1.0cm]{Alice};
\node (chan) at (cchan) [rectangle, draw, minimum width=2.0cm, minimum height=1.5cm]{$p(y_2, y_3,z|x_1,x_2)$};
\node (dec1) at (cdec1) [rectangle, draw, minimum height=1.2cm, minimum width=1.0cm]{Bob};
\node (dec2) at (cdec2) [rectangle, draw, minimum height=1.2cm, minimum width=1.0cm]{Eve};
\node (rel) at (ck1) [rectangle, draw, minimum height=1.0cm]{$Rel$};

\draw[->] (enc) -- node[above]{$\mathbf X_1$} (chan);
\draw[->] (chan.north east) ++(0,-0.2) --node[above]{$\mathbf Y_3$} ++(1,0);
\draw[->] (chan.south east) ++(0,0.2) --node[above] {$\mathbf Z$} ++(1,0);

\begin{scope}[dashed]
\end{scope}

\draw[<-] (enc.160) --++(-0.5,0) node[left]{$W$};

\draw[->] (dec1.20) --++(0.4,0) node[right]{$W$};
\draw[->] (dec2.20) --++(0.4,0) node[right]{$\xcancel{W}$};
\draw[->](2.8,0.9)--(2.8,2.3);
\draw[<-](3.2,1)--(3.2,2.3);
\node[above] at (3.5,1.3){$\mathbf{X}_2$};
\node[above] at (2.5,1.3){$\mathbf{Y}_2$};
\node[right] at (3.5,3.0){$\mathbf {\widehat Y}_2$};
\end{tikzpicture}
\caption{Relay-Eavesdropper channel with trusted relay and a confidential message intended for Bob. RV~$X_2$: relay's input to the channel. RV~$Y_2$: the channel's output as seen by the relay}
\label{fig:rel_eve_tr_weakrel}
\end{center}
\end{figure}
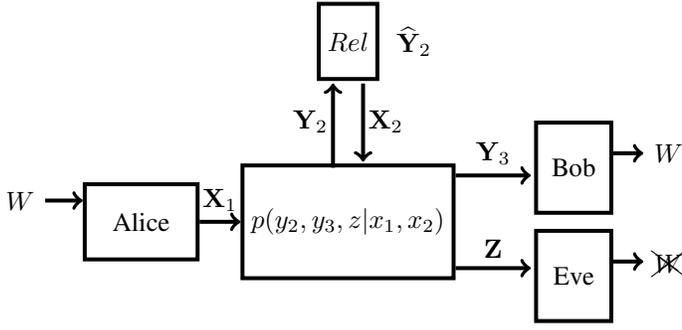

\noindent The notation is standard, but we describe it here for completeness. The system model is depicted in Fig.~\ref{fig:rel_eve_tr_weakrel}. We assume a two receiver discrete memoryless relay-eavesdropper channel with a confidential message intended for one of the receivers, with the other acting as an eavesdropper. The finite sets $\mathcal X_1, \mathcal X_2$, $\mathcal Y_2,\mathcal Y_3, \mathcal Z$ respectively represent the channel's input at node~$1$ (the transmitter), at node~$2$ (the relay), the channel's output at nodes~$2$, $3$ (legitimate receiver aka Bob) and $4$ (eavesdropper aka Eve). Finally, $\mathcal {\hat Y}_2$ represents the alphabet of the compression random variable. The channel is described by the conditional probability distribution $P_{Y_2,Y_3,Z|X_1,X_2}$, where RVs $X_i \in \mathcal X_i,\,i=1,2$ and $Y_i \in \mathcal Y_i,\,i=2,3$ and $Z \in \mathcal Z$. The transmitter intends to send an independent message $W \in \{1,2,\dots,2^{nR_1}\} \overset{\Delta}= \mathcal W$ to the receiver Rx~$Y_3$ in $n$ channel uses while ensuring information theoretic secrecy, defined below. The channel is memoryless and without feedback i.e. $\forall (\mathbf x_1, \mathbf x_2) \in \prod_{t=1}^2\mathcal X_t^n, \mathbf y_t \in \mathcal Y_t^n,\,t=2,3, \mathbf z \in \mathcal Z^n$,
$\begin{aligned}[t]
 P(\mathbf y_2, \mathbf y_3, \mathbf z|\mathbf x_1, \mathbf x_2) = \prod_{i=1}^n P_{Y_2,Y_3,Z|X_1,X_2}(y_{2i},y_{3i},z_{i}|x_{1i},x_{2i})
\end{aligned}$

\noindent The decoding function at Bob $\equiv Y_3$ is a map $\phi_3: \mathcal Y_3^n \times \mathcal Y_3^n \rightarrow \mathcal W \times \widetilde {\mathcal W}$.


\noindent A $(2^{nR_1}, n, P_e^{(n)})$ code for the relay-eavesdropper channel with compress-forward (CF) consists of a (stochastic) encoding function at the Tx and an encoding function at the relay, and a decoding function $\phi_3$ at the destination and error probability $\begin{aligned}[t] P_e^{(n)} = \sum_{(w, \tilde w)} \frac{\Pr[\phi_3(\mathbf Y_3^n \times \mathbf Y_3^n) \neq (w, \tilde w)|(w, \tilde w) ]}{2^{nR_1} \times 2^{n\tilde R_1}}\end{aligned}$

\noindent A secrecy rate $R_1$ is said to be achievable for the DM relay-eavesdropper channel if, for any $\epsilon > 0$, $\exists (2^{n[R_1+\tilde R_1]}, n, P_e^{(n)})$ code s.t. the following requirements are satisfied:
\begin{align*}
 &\text{reliability: } P_e^{(n)} \leq \epsilon \\
 &\text{(weak) secrecy: } H(W^{[B-1]}|\mathbf Z^{[B]}) \geq n(B-1)R_1 - n(B-1)\epsilon
\end{align*}
We use the notation $[x]^+ \overset{\text{def}}= \max\{x,0\}$.
\section{Inner Bound}
\label{sec:rel_eve_cf_ib}
\noindent The main result of our paper is presented below, namely, that judicious independent choice of the compression sequence rate $\hat R$ and WZ binning rate $\hat R - R_2$,\footnote{Equivalently, of $\hat R$ and $R_2$} enables us to maximize the secrecy rate. 

%
%
\begin{theorem}
\noindent We first define:
\begin{align*}
 WZ^{Bob} \overset{\text{def}}{=} I(\hat Y_2;X_1,Y_3|X_2),\,
 WZ^{Eve} \overset{\text{def}}{=} I(\hat Y_2;X_1,Z|X_2) 
\end{align*}

\noindent A (pure secrecy) rate $R_1$ is achievable if there exists distributions $P_{X_1}P_{X_2}P_{\hat Y_2|X_2,Y_2}$ s.t. the following hold: \\

\noindent \fbox{Case~$1$}: $I(X_2;Z) + WZ^{Eve} < I(X_2;Y_3) + WZ^{Bob}$.
 
\noindent In all Case~$1$ sub-cases, we choose $\hat R \in (I(X_2;Z) + WZ^{Eve}, I(X_2;Y_3) + WZ^{Bob})$ and $R_2 > I(X_2;Z_3)$. So Eve cannot uniquely decode $\mathbf x_2$, and either decodes it nonuniquely or treats it as noise, depending on the sub-cases.

\noindent \fbox{Case~$1(a)$}: $
I(X_2;Z) + WZ^{Eve} < I(X_2;Y_3) + I(\hat Y_2;Y_3|X_2)$
            
\noindent \fbox{Case~$1(a)(i)$}: $I(X_2;Z|X_1) < I(X_2;Y_3) + I(\hat Y_2;Y_3|X_2)$. The choices $R_2 = \max\{I(X_2;Y_3), I(X_2;Z|X_1)\}+\epsilon$ and $\hat R = I(X_2;Y_3)+I(\hat Y_2;Y_3|X_2)-\epsilon > I(X_2;Z)+WZ^{Eve}$, enable the secrecy rate $R_1 = [I(X_1; \hat Y_2,Y_3|X_2) - I(X_1;Z)]^+$.
             
\noindent \fbox{Case~$1(a)(ii)$}: $I(X_2;Y_3) + I(\hat Y_2;Y_3|X_2) < I(X_2;Z|X_1)$. The choices $R_2 = \max\{I(X_2;Y_3), I(X_2;Z)\}+\epsilon$, $\hat R$ as before, enable
 $R_1 = \Big[I(X_1; \hat Y_2,Y_3|X_2) - [I(X_1,X_2;Z) - R_2] \Big]^+$. \textit{The penalty term can be reduced by choosing $\epsilon>0$ s.t. $R_2 \uparrow \hat R$}.
            
\noindent \fbox{Case~$1(b)$}: $I(X_2;Y_3) + I(\hat Y_2;Y_3|X_2) < I(X_2;Z) + WZ^{Eve} < I(X_2;Y_3) + WZ^{Bob}$
        

\noindent \fbox{Case~$1(b)(i)$}: $I(X_2;Z|X_1) < I(X_2;Y_3) + WZ^{Bob}$. The choices $R_2 = \max\{I(X_2;Z|X_1), I(X_2;Y_3)\}+\epsilon (> I(X_2;Z))$ and $\hat R \in (I(X_2;Z)+WZ^{Eve}, I(X_2;Y_3)+WZ^{Bob})$ 
enable the secrecy rate $R_1 = \Big[I(X_1;\hat Y_2, Y_3|X_2) + I(X_2;Y_3) + I(\hat Y_2;Y_3|X_2) - \hat R - I(X_1;Z) \Big]^+$. By choosing $\hat R \downarrow \max\{I(X_2;Z|X_1), I(X_2;Y_3)\}$, we can obtain rate $R_1 =
\Big[I(X_1;\hat Y_2, Y_3|X_2) + I(X_2;Y_3) + I(\hat Y_2;Y_3|X_2) - \max\{I(X_2;Z|X_1), I(X_2;Y_3)\} - I(X_1;Z) \Big]^+$.

\noindent \fbox{Case~$1(b)(ii)$}: $I(X_2;Y_3) + WZ^{Bob} < I(X_2;Z|X_1)$. The choices $R_2 > \max\{I(X_2;Y_3), I(X_2;Z)\}+ \epsilon$ and $\hat R \in (I(X_2;Z)+WZ^{Eve}, I(X_2;Y_3)+WZ^{Bob})$ enable the secrecy rate $R_1 = \Big[I(X_1;\hat Y_2, Y_3|X_2) + I(X_2;Y_3)  + I(\hat Y_2;Y_3|X_2) - \hat R - [I(X_1,X_2;Z) - R_2] \Big]^+$. Choosing $\hat R \downarrow I(X_2;Z)+WZ^{Eve},\,R_2 \uparrow \hat R$ enables rate $R_1 = \Big[I(X_1;\hat Y_2, Y_3|X_2) + I(X_2;Y_3)  + I(\hat Y_2;Y_3|X_2) - I(X_1,X_2;Z) \Big]^+$
 
\noindent \fbox{Case~$2$}: $I(X_2:Y_3) + WZ^{Bob} < I(X_2;Z) + WZ^{Eve}$
       
\noindent \fbox{Case~$2(a)$}:  $WZ^{Eve} < WZ^{Bob} (\implies I(X_2;Y_3) < I(X_2;Z))$
              
\noindent \fbox{Case~$2(a)(i)$}: $WZ^{Eve} < I(\hat Y_2;Y_3|X_2)$. The choices $R_2 \in (I(X_2;Y_3), I(X_2;Z)),\, \hat R - R_2 > WZ^{Eve}, \hat R < I(X_2;Y_3) + I(\hat Y_2;Y_3|X_2)$ can be made consistent and enable secrecy rate $R_1 = [I(X_1; \hat Y_2,Y_3|X_2) - I(X_1;Z|X_2)]^+$.
               
\noindent \fbox{Case~$2(a)(ii)$}: $I(\hat Y_2;Y_3|X_2) < WZ^{Eve} < WZ^{Bob}$. Choosing $R_2 > I(X_2;Y_3), \hat R - R_2 > WZ^{Eve}, \hat R < I(X_2;Y_3)+WZ^{Bob}$ enables secrecy rate $R_1 = \Big[[I(X_1;Y_3|X_2) - I(X_1;Z|X_2)] + [WZ^{Bob} - WZ^{Eve}]\Big]^+$.
               
\noindent \fbox{Case~$2(b)$}:  $WZ^{Bob} < WZ^{Eve}$

\noindent \fbox{Case~$2(b)(i)$}: $I(X_2;Y_3) < I(X_2;Z)$. Choosing $R_2 < I(X_2;Y_3),\,\hat R > I(X_2;Z)+WZ^{Eve} \implies \hat R - R_2 > WZ^{Eve}$ and hence also $\hat R - R_2 > WZ^{Bob}$, which enables the secrecy rate $R_1 = [I(X_1;Y_3|X_2) - I(X_1;Z|X_2)]^+$.
               
\noindent \fbox{Case~$2(b)(ii): I(X_2;Z) < I(X_2;Y_3)$}. \\
\noindent \fbox{$(A)$}: $I(X_2;Z) < I(X_2;Y_3) < I(X_2;Z|X_1)$. The choices $R_2 \in (I(X_2;Z),I(X_2;Y_3)),\,\hat R - R_2 > WZ^{Eve}$ enables secrecy rate $R_1 = \Big[I(X_1;Y_3|X_2) - [I(X_1,X_2;Z) - R_2]\Big]^+$. \textit{The penalty term is reduced by choosing $R_2 \uparrow I(X_2;Y_3)$}
                      
\noindent \fbox{$(B)$}: $I(X_2;Z) < I(X_2;Z|X_1) < I(X_2;Y_3)$. The choices $R_2 \in (I(X_2;Z|X_1),I(X_2;Y_3))$ and $\hat R - R_2$ as before, enables secrecy rate $R_1 = [I(X_1;Y_3|X_2) - I(X_1;Z)]^+$.
                     
\end{theorem}

\section{Achievability Scheme}
\label{sec:rel_eve_cf_ach}

\noindent As is common in secrecy scenarios, the transmitter's codebook is divided into bins  (aka subcodebooks), with the bin size intended to confuse Eve. We use \cite{LuoGohYanik2012}'s  decoding technique. \\

\noindent \textit{Codebook Generation: } There are $B$ blocks of transmission. An independent set of codebooks is generated for each block, and these are known to all parties involved. Aside from an additional binning structure, the codebook construction at the transmitter is almost identical to that in the canonical relay channel \cite{ElGamKim2012, CovElGam79}. The transmitter codebook in block~$b$ is given by $\begin{aligned}[t] \mathcal C_1^{(b)} = \{\mathbf x_1^{(b)}(w,\tilde w)|w \in [2^{nR_1}], \tilde w \in [2^{n\tilde R_1}])\}\end{aligned}$, with the codewords generated independently using distribution $P_{X_1}$ i.e. $\begin{aligned}[t] \mathbf x_1^{(b)}(w,\tilde  w) \sim \prod_{i=1}^nP_{X_1}(x_{1i}^{(b)})\end{aligned}$. For a fixed $w=1$ (say), the set $\mathbf x_1(1,\tilde w), \tilde w \in [2^{n\tilde R_1}]$ can be thought of as a bin. 
The relay channel codebook $\mathcal C_2^{(b)}$ and compression codebook associated with each relay channel codeword is exactly identical to that in \cite{ElGamKim2012, CovElGam79}. $|\mathcal C_2^{(b)}| = 2^{nR_2}$. The compression codebook consists of $2^{nR_2}$ --  i.e., one per relay codeword -- satellite codebooks; each is of size $2^{n\hat R}$, and is divided into $2^{nR_2}$ WZ bins of size $2^{n[\hat R - R_2]}$. \textit{The generation of the codewords is identical to that in \cite{ ElGamKim2012, CovElGam79} and is omitted due to space}. See \cite{KIyer2018b} for details. 
\begin{remark}
There are two differences with the codebooks used for NF and GNF in \cite{LaiElGam08}. In \cite{LaiElGam08}, the relay does not WZ bin the compression sequences $\mathbf {\hat y}_2$. Further, each $\mathbf {\hat y}_2$ is associated with a set of different possible relay codewords $\mathbf x_2$ -- this is called NF. Once $\mathbf {\hat y}_2$ is known, the choice of $\mathbf x_2$ to transmit is determined by the relay's private randomness. In our scheme, on the other hand, the relay does WZ bin the $\mathbf {\hat y}_2$s, and the $\mathbf x_2$ transmitted is wholly determined by the WZ bin of the previous block's $\mathbf {\hat y}_2$ -- private randomness not needed. This determinism is crucial for the success of the equivocation calculations. Since this gives a greater degree of control to the transmitter, we also have reason to believe that this could lead to an improvement in achievable secrecy rate.
\end{remark}
\noindent \textit{Transmitter Encoding:} To send $W_j \in [1:2^{nR_1}]$ in a block $b \in {1,2,\dots,B-1}$, the transmitter chooses uniformly at random from among $2^{n\tilde R_1}$ possibilities for $\tilde w$ inside the corresponding bin $W_j$. The corresponding codeword $\mathbf x_1$ is transmitted. \\
\noindent \textit{Relay Encoding:} The relay encoding is standard \cite{ElGamKim2012, CovElGam79} i.e. it looks for a compression codeword that satisfies the joint typicality condition $\big( \mathbf x_2, \mathbf {\hat y}_2, \mathbf y_2 \big) \in \mathcal T^{(n)}_\epsilon$ and then determines $\mathbf {\hat y}_2$'s WZ bin index. The relay codeword corresponding to this WZ bin is transmitted in the next block. \\
\noindent \textit{Decoding at Bob: } We use the decoding techniques developed by \cite{LuoGohYanik2012}. See \cite{KIyer2018b}.
These give rise to:
\begin{subequations}
\begin{align}
\hat R \! &< I(X_2;Y_3) + I(\hat Y_2;X_1,Y_3|X_2) \label{eq:bob_dec_1a}\\ 
 R_1 \! + \! \tilde R_1 \!  &< I(X_1;\! \hat Y_2,\!Y_3|X_2) \label{eq:bob_dec_1b}\\
 R_1 \! + \! \tilde R_1 \!  &<  I(X_1;\! \hat Y_2,\!Y_3|X_2) \notag \\
                     &\phantom{wwww} + [ I(\hat Y_2;Y_3|X_2) + I(X_2;Y_3) - \hat R] \label{eq:bob_dec_1c}\\
 R_1 \! + \! \tilde R_1 \! &< I(X_1;\! \hat Y_2,Y_3|X_2) \notag \\
                    &\phantom{wwww} + [ I(\hat Y_2;Y_3|X_2) +  \phantom{ww} R_2 - \hat R] \label{eq:bob_dec_1d}
\end{align}
\end{subequations}
\begin{remark}
Consider the canonical ``weak'' relay with no secrecy requirement i.e., $\tilde R_1 = 0$, as in \cite{LuoGohYanik2012}. We observe that: 
\begin{itemize}[leftmargin=*,nolistsep]
 \item If $\hat R \overset{\alpha > 0}{=} I(X_2;Y_3) + I(\hat Y_2;Y_3|X_2) + \alpha$, then $R_1  < I(X_1; \hat Y_2,Y_3|X_2) - \alpha$ i.e. too many compression sequences reduce achievable rate $R_1$. (\cite{LuoGohYanik2012}, \cite{WuFanXie10}).
 
 \item If $\hat R < I(X_2;Y_3) + I(\hat Y_2;Y_3|X_2)$, if in addition $R_2 \overset{\beta > 0}{=} \hat R - I(\hat Y_2;Y_3|X_2) - \beta$, then again $R_1 < I(X_1; \hat Y_2,Y_3|X_2) - \beta$ i.e. too few relay codewords also reduce achievable rate.  
 
 \item If $\hat R - R_2 > WZ^{Bob}$, then Bob cannot uniquely decode $\mathbf {\hat y}_2$.
\end{itemize}
\end{remark}

\begin{remark} 
  In the ``weak'' relay-eavesdropper channel, $\tilde R_1 \neq 0$; $\hat R$ \& $\hat R - R_2$ chosen to maximize secrecy rate: $R_1$ 
\end{remark}
\begin{remark}
If either $\eqref{eq:bob_dec_1a}$ (Bob's proxy constraint) or $ R_2 < I(X_2;Y_3)$ is satisfied, then Bob can decode $\mathbf x_2$ uniquely. In Cases~$1(a)(i), 1(a)(ii), 2(a)(i)$, $\because R_2 > I(X_2;Y_3), \hat R < I(X_2;Y_3) + I(\hat Y_2;Y_3|X_2)$, of the three constraints on $R_1 + \tilde R_1$, the tightest constraint is $\eqref{eq:bob_dec_1b}$. In Cases~$1(b)(i), 1(b)(ii), 2(a)(ii)$, $\because R_2 > I(X_2;Y_3), \hat R > I(X_2;Y_3) + I(\hat Y_2;Y_3|X_2)$, of the three constraints on $R_1 + \tilde R_1$, the tightest constraint is $\eqref{eq:bob_dec_1c}$. In Cases~$2(b)(i), 2(b)(ii)$, constraint $\eqref{eq:bob_dec_1a}$ is violated, but $R_2 < I(X_2;Y_3)$ still holds. So Bob uses $(\mathbf x_2, \mathbf y_3) \in \mathcal T^{(n)}_\epsilon$ i.e., single block information to decode $\mathbf x_2$, and then conditions on $\mathbf x_2$ to decode $\mathbf x_1$ using $(\mathbf x_1, \mathbf x_2, \mathbf y_3) \in \mathcal T^{(n)}_\epsilon$ at rate $R_1 + \tilde R_1 \approx I(X_1;Y_3|X_2)$. \textit{It can be s.t. decoding $\mathbf {\hat y}_2$ nonuniquely would be suboptimal -- details omitted}. 
\end{remark}
\begin{remark}
 In Case~$1$ i.e., $I(X_2;Z)+WZ^{Eve} < I(X_2;Y_3)+WZ^{Bob}$, we will always choose $\hat R \in ( I(X_2;Z)+WZ^{Eve}, I(X_2;Y_3)+WZ^{Bob})$, ensuring that Eve's proxy constraint, to wit, $\hat R < I(X_2;Z)+WZ^{Eve}$ is violated, and by additionally choosing  $R_2 > I(X_2;Z)$, Eve cannot decode $\mathbf x_2$ uniquely and consequently cannot decode $\mathbf {\hat y}_2$ either. 
 \end{remark}

\begin{remark}
 In Case~$1$ scenarios, the $\mathbf x_2$ in the first block is known to all parties, including Eve. So, in the first block, Eve can decode $\mathbf x_1$ at a rate $\tilde R_1 \approx I(X_1;Z|X_2)$. However, this holds only for the first block, creating a boundary effect that does not affect the overall secrecy rate achievable. 
\end{remark}
\begin{remark}
 In all Case~$2$ scenarios i.e., $I(X_2;Y_3)+WZ^{Bob} < I(X_2;Z)+WZ^{Eve}$, we will choose $\hat R - R_2 > WZ^{Eve}$, thus ensuring that Eve cannot decode $\mathbf {\hat y}_2$ uniquely. But in Case~$2a(i),(ii),2b(i), R_2 < I(X_2;Z)$, enabling Eve to decode $\mathbf x_2$ uniquely. Consequently, in these cases only, Eve can decode $\mathbf x_1$ at rate $\tilde R_1 \approx I(X_1;Z|X_2)$.
\end{remark}
\begin{remark}
In all other cases, whether Eve decodes $\mathbf x_2$ nonuniquely or treats $\mathbf x_2$ as noise will be determined by the exact choices of $\hat R$ and $\hat R - R_2$; the choices in turn depend on the sub-case. Now there are the following possibilities: 
\begin{enumerate}[leftmargin=*,nolistsep]
 \item $R_2 \in (I(X_2;Z), I(X_2;Z|X_1))$. Given the message $W$, Eve decodes $\mathbf x_2$ nonuniquely and decodes $\mathbf x_1$ uniquely at rate $\tilde R_1 = I(X_1,X_2;Z) - R_2$.
 \item $R_2 > I(X_2;Z|X_1)$. Given the message $W$, treat $\mathbf x_2$ as noise and decode $\mathbf x_1$ at rate $\tilde R_1 = I(X_1;Z)$
\end{enumerate}
\end{remark}

\section{Equivocation Calculations}
\label{sec:equiv_cal}
We have separate equivocation calculations for Cases~$1$, $2(a), 2(b)(ii)$. Calculation for $2(b)(i)$ is identical to Case~$2(a)$. 

\noindent \fbox{\textbf{Case~$1$: $I(X_2;Z)+WZ^{Eve} < I(X_2;Y_3)+WZ^{Bob}$}}
In this regime, we will always choose $\hat R \in (I(X_2;Z)+WZ^{Eve}, I(X_2;Y_3)+WZ^{Bob})$, and 
$R_2 > I(X_2;Z)$. 
\begin{figure*}[h]
\hrule Equivocation calculation for \fbox{\textbf{Case~$1$: $I(X_2;Z)+WZ^{Eve} < I(X_2;Y_3)+WZ^{Bob}$}}
\begin{align}
  H(W^{[B-1]}|\mathbf Z^{[B]}) 
  &\geq I(\mathbf X_1^{[B]};W^{[B-1]}|\mathbf Z^{[B]}) 
  = \underset{(i)}{H(\mathbf X_1^{[B]}|\mathbf Z^{[B]})} - \underset{(ii) \approx (B-1)n\epsilon \text{ (see main text)}}{H(\mathbf X_1^{[B]}|W^{[B-1]},\mathbf Z^{[B]})} \label{eq:equivcal1:a} \\
\text{Term} \fbox{\eqref{eq:equivcal1:a}(i)}: \text{ i.e. } H(\mathbf X_1^{[B]}|\mathbf Z^{[B]}) &= 
\underset{(i)}{H((\mathbf X_1, \mathbf X_2, \mathbf {\hat Y}_2, \mathbf Z)^{[B]})}
- \underset{(ii)}{H((\mathbf X_2, \mathbf {\hat Y}_2)^{[B]}|(\mathbf X_1,\mathbf Z)^{[B]})} - \underset{(iii)}{H(\mathbf Z^{[B]})} \label{eq:equivcal1:aa}\\
 \text{Term }\fbox{\eqref{eq:equivcal1:aa}(i)}: \text{ i.e. } H((\mathbf X_1, \mathbf X_2, \mathbf {\hat Y}_2, \mathbf Z)^{[B]}) &= 
 \sum_{j=1}^B\underset{T_j}{H((\mathbf X_1, \mathbf X_2, \mathbf {\hat Y}_2,
 \mathbf Z)_j|(\mathbf X_1, \mathbf X_2, \mathbf {\hat Y}_2, \mathbf Z)^{[j-1]})} 
\label{eq:equivcal1:aaa} 
\end{align}
\noindent
\begin{align}
  &\text{In }\eqref{eq:equivcal1:aaa},\, T_j = \underset{(a)}{H(\mathbf X_{2,j}|(\mathbf X_1, \mathbf X_2, \mathbf {\hat Y}_2, \mathbf Z)^{[j-1]})} + \underset{(b)}{H(\mathbf X_{1,j}|\mathbf X_{2,j},(\mathbf X_1, \mathbf X_2, \mathbf {\hat Y}_2, \mathbf Z)^{[j-1]})} \notag\\
  &\phantom{wwwwwwwwwwwww} + \underset{(c)}{H(\mathbf Z_j|(\mathbf X_1, \mathbf X_2)_j, (\mathbf X_1, \mathbf X_2, \mathbf {\hat Y}_2, \mathbf Z)^{[j-1]})}
   + \underset{(d)}{H(\mathbf {\hat Y}_{2,j}|(\mathbf X_1, \mathbf X_2, \mathbf Z)_j , (\mathbf X_1, \mathbf X_2, \mathbf {\hat Y}_2, \mathbf Z)^{[j-1]})}\label{eq:equivcal1:aaaa} \\
&\text{Terms } \fbox{\eqref{eq:equivcal1:aaaa}(a)} = 0,
\fbox{\eqref{eq:equivcal1:aaaa}(b)}= n[R_1 + \tilde R_1],
\fbox{\eqref{eq:equivcal1:aaaa}(c)}= H(\mathbf Z_j|\mathbf X_{1,j}, \mathbf X_{2,j}) \label{eq:equivcal1:b}, \fbox{\eqref{eq:equivcal1:aaaa}(d)}=n[\hat R - I(\hat Y_2;X_1,Z|X_2)] \\
&\text{(Foregoing evaluated in main text). We can write: }\phantom{www} 
T_j = n[R_1 + \tilde R_1 + H(\mathbf Z_j|\mathbf X_{1,j}, \mathbf X_{2,j}) + \hat R - I(\hat Y_2;X_1,Z|X_2)] 
\end{align}
\begin{align}
&\text{Term } \fbox{\eqref{eq:equivcal1:aa}(ii)} \text{ i.e. }H((\mathbf X_2, \mathbf {\hat Y}_2)^{[B]}|(\mathbf X_1,\mathbf Z)^{[B]}) =  \sum_{j=1}^B\underset{\tilde T_j}{H((\mathbf X_2, \mathbf {\hat Y}_2)_j|(\mathbf X_2, \mathbf {\hat Y}_2)^{[j-1]}, \mathbf X_1^{[B]},\mathbf Z^{[B]})} \label{eq:equivcal1:aaaaa}\\
&\text{In } \eqref{eq:equivcal1:aaaaa},\,\tilde T_j = \underset{(=0, \text{ for the same reason as Term } \fbox{\eqref{eq:equivcal1:aaaa}(a)})}{H(\mathbf X_{2,j}|(\mathbf X_2, \mathbf {\hat Y}_2)^{[j-1]}, (\mathbf X_1,\mathbf Z)^{[B]})} + \underset{(\tilde b) \text{ (evaluated in main text) }}{H(\mathbf {\hat Y}_{2,j}|\mathbf X_{2,j}, (\mathbf X_2, \mathbf {\hat Y}_2)^{[j-1]}, (\mathbf X_1,\mathbf Z)^{[B]})}  \label{eq:equivcal1:aaaaaa} \\
&\phantom{wwwwww}=n[\hat R - \min\{R_2,I(X_2;Z|X_1)\}- I(\hat Y_2;X_1,Z|X_2)] \\
 &\text{Term }\fbox{\eqref{eq:equivcal1:aa}(iii)} \text{ i.e. } H(\mathbf Z^{[B]}) = \sum_{j=1}^BH(\mathbf Z_j|\mathbf Z^{[j-1]}) \leq \sum_{j=1}^BH(\mathbf Z_j) \label{eq:equivcal1:aaaaaaa}
\end{align}
Consider an individual block~$j=1,2,\dots,B-1$. We can lower bound 
\begin{align}
 T_j - \tilde T_j - H(\mathbf Z_j|\mathbf Z^{[j-1]}) 
 \geq T_j - \tilde T_j - H(\mathbf Z_j) 
 =n[R_1 + \tilde R_1 +\min\{R_2, I(X_2;Z|X_1)\} - I(\mathbf X_{1,j}, \mathbf X_{2,j}; \mathbf Z_j)]  \label{eq:equivcal1:aaaaaaaa}
\end{align}
\begin{align}
 \text{From } \eqref{eq:equivcal1:aa}, \text{ we have } H(\mathbf X_1^{[B]}|\mathbf Z^{[B]}) 
 &= \fbox{\eqref{eq:equivcal1:aa}(i)} - \fbox{\eqref{eq:equivcal1:aa}(ii)} - \fbox{\eqref{eq:equivcal1:aa}(iii)} \geq \sum_{j=1}^B [T_j - \tilde T_j - H(\mathbf Z_j)] \\
 \text{Above and } \eqref{eq:equivcal1:a} \text{ give }H(\mathbf W^{[B-1]}|\mathbf Z^{[B]}) 
 &\geq \sum_{j=1}^{B-1} [T_j - \tilde T_j - H(\mathbf Z_j)] - H(\mathbf X_1^{[B]}|W^{[B-1]},\mathbf Z^{[B]}) \label{eq:equivcal1:aaaaaaaaa}
\end{align}
where in the summation above, we have replaced $B \leftarrow B-1$, using the fact that since no new message is transmitted in block~$B$, $\mathbf X_{1,B}$ is fixed beforehand and known to all parties, and so the last block's contribution is zero.
\newline 
\hrule
\end{figure*}
%
\noindent
We s.t. $\fbox{\text{Term }\eqref{eq:equivcal1:a}(ii)}$ ``small''. The rate of a transmitter bin is $\tilde R_1 \approx I(X_1;Z)\mathbb I\{R_2 > I(X_2;Z|X_2)\} + [I(X_1,X_2;Z) - R_2]\mathbb I\{R_2 \in (I(X_2;Z), I(X_2;Z|X_1))\}$. 
Conditioning on transmitted $W_j$ is equivalent to knowing the bin and 
reduces the number of possibilities for $\mathbf X_{1,j}$ from $2^{n[R_1 + \tilde R_1]}$ to $2^{n\tilde R_1}$, and knowing $\mathbf Z_j$, enables us to decode $\mathbf X_{1,j}$ inside the bin correctly whp, using, in every block $j=1,2,\dots,B-1$:
\begin{itemize}[leftmargin=*,nolistsep]
 \item either $(\mathbf x_{1,j}, \mathbf z(j)) \in \mathcal T^{(n)}_\epsilon$ i.e.  treating $\mathbf x_{2,j}$ as noise if $R_2 > I(X_2;Z|X_1)$.
 
 \item or $(\mathbf x_{1,j}, \mathbf x_{2,j},\mathbf z(j)) \in \mathcal T^{(n)}_\epsilon$ for a unique $\mathbf x_1$ and some $\mathbf x_2$ at a rate $I(X_1,X_2;Z) - R_2$ if $R_2 \in (I(X_2;Z), I(X_2;Z|X_1))$.
\end{itemize} 
In the last block no new message is transmitted and thus $\mathbf X_{1,B}$ is known. A standard application of Fano's inequality gives: 
$ \fbox{\text{Term }\eqref{eq:equivcal1:a}(ii)} \leq (B-1) \times n\epsilon$.\\

\noindent
\text{Term } \fbox{$\eqref{eq:equivcal1:aaaa}(a)$} = 0. No noise-forwarding i.e.  $\mathbf X_{2j}$ \text{is a function of} $(\mathbf X_{2,j-1}, \mathbf {\hat Y}_{2,j-1})$ present in the conditioning. 

\noindent 
Term \fbox{$\eqref{eq:equivcal1:aaaa}(b)$}= $n[R_1 + \tilde R_1]$. $\mathbf X_{1,j}$ is uniformly chosen from $2^{n[R_1 + \tilde R_1]}$ codewords; choice unaffected by the conditioning.

\noindent 
Term \fbox{$\eqref{eq:equivcal1:aaaa}(c)$} $=H(\mathbf Z_j|\mathbf X_{1,j}, \mathbf X_{2,j})$. $\because$ Channel is DMC, \\
$\mathbf Z_j \leftrightarrow (\mathbf X_{1,j}, \mathbf X_{2,j}) \leftrightarrow (\mathbf X_1, \mathbf X_2, \mathbf {\hat Y}_2, \mathbf Z)^{[j-1]}$ are Markov RVs. 

\noindent
$\begin{aligned}[t]\text{Term }\fbox{\eqref{eq:equivcal1:aaaa}(d)} = n[\hat R - I(\hat Y_2; X_1,Z|X_2)]\end{aligned}$. The first three quantities in the conditioning, namely, $(\mathbf X_1, \mathbf X_2, \mathbf Z)_j$ can be used to reduce the number of possibilities for $\mathbf {\hat y}_{2,j}$ as follows. Knowledge of $\mathbf X_{2,j}$ reduces the number of possibilities for the compression sequence down to $2^{n\hat R}$. Now by the condition $\begin{aligned}[t]
 \big( \mathbf x_{1,j}, \mathbf x_{2,j}, \mathbf {\hat y}_2(|\mathbf x_{2,j}), \mathbf z_j\big) \in \mathcal T^{(n)}_\epsilon
\end{aligned}$, we can create an equiprobable list of size $2^{n[\hat R - I(\hat Y_2; X_1,Z|X_2)]}$. \\
\noindent
Term \fbox{$\eqref{eq:equivcal1:aaaaaa}(\tilde b)$.} Knowledge of $\mathbf X_{2,j}$ reduces the possibilities for $\mathbf {\hat Y}_{2,j}$ down to $2^{n\hat R}$. Given $(\mathbf X_{1,j+1}, \mathbf Z_{j+1})$, we can further reduce the number of possible  codewords $\mathbf X_{2,j+1}$ in block~$j+1$ down to $2^{n[R_2 - I(X_2;Z|X_1)]}$. Each such codeword corresponds to a separate WZ bin, each of size $2^{n[\hat R - R_2]}$. Now the size of the list of possible $\mathbf {\hat y}_2$ is $\begin{aligned}[t]
 2^{n[R_2 - I(X_2;Z|X_1)]} \times 2^{ n[\hat R - R_2]}
\end{aligned}$ where the first term will be $1$ if $R_2 < I(X_2;Z|X_1)$. In terms of rate, we have $\begin{aligned}[t]
 n[\cancel{R_2} - \min\{R_2, I(X_2;Z|X_1)\}] + n[\hat R - \cancel{R_2}]
\end{aligned}$. Conditioning on $(\mathbf X_{2,j}, \mathbf X_{1,j}, \mathbf Z_j)$ gives us a further reduction by: $2^{nI(\hat Y_2; X_1,Z|X_2)}$ so that, finally, the 
Term \\ \fbox{$\eqref{eq:equivcal1:aaaaaa}(\tilde b)$}
$\begin{aligned}[t]
=n[\hat R \!-\! \min\{R_2,\! I(X_2;Z|X_1)\}\! -\! I(\hat Y_2; X_1,Z|X_2)]
\end{aligned}$.\\

\noindent
We now lower bound the RHS of \fbox{$\eqref{eq:equivcal1:aaaaaaaa}$} by first computing two upper bounds for $I(\mathbf X_{1,j}, \mathbf X_{2,j}; \mathbf Z_j)$. We have (see  \cite{TLSP2011} and \cite{LMSY08}) 
\begin{align}
 I(\mathbf X_{1,j}, \mathbf X_{2,j}; \mathbf Z_j) &\leq nI(X_1,X_2;Z) + n\epsilon \,\,\text{(u.b. I)} \notag \\ 
 I(\mathbf X_{1,j}, \mathbf X_{2,j}; \mathbf Z_j) &= I(\mathbf X_{2j}; \mathbf Z_j) + I(\mathbf X_{1,j}; \mathbf Z_j|\mathbf X_{2,j}) \notag \\
 &\leq H(\mathbf X_{2,j}) + nI(X_1;Z|X_2) + n\epsilon \\
 &\leq nR_2 + nI(X_1;Z|X_2) + n\epsilon \,\, \notag \text{(u.b. II)} \implies \\
I(\mathbf X_{1,j},\! \mathbf X_{2,j};\! \mathbf Z_j)\! &\leq\! n[\min\{R_2, I(X_2;Z)\}\! +\! I(X_1;Z|X_2)\! + \epsilon] \label{eq:equivcal1:mt}
\end{align}

\noindent We use the above to obtain a lower bound for the RHS of $\eqref{eq:equivcal1:aaaaaaaa}$ to obtain, for a single block:
\begin{align*}
 &\phantom{w}nR_1 + n\tilde R_1 + n\min\{R_2, I(X_2;Z|X_1)\}] \\ 
 &- n\min\{R_2, I(X_2;Z)\} - nI(X_1;Z|X_2) - n\epsilon
\end{align*}
\begin{itemize}
\item If $R_2 > I(X_2;Z|X_1) > I(X_2;Z)$, then $\tilde R_1 \approx I(X_1;Z)$, and the above becomes 
$nR_1 - n\epsilon$.
%
 \item If $R_2 \in (I(X_2;Z), I(X_2;Z|X_1))$, then $\tilde R_1 = I(X_1,X_2;Z)-R_2$, and we again have $nR_1 - n\epsilon$.
 %
\end{itemize}

\noindent Using this lower bound for RHS of $\eqref{eq:equivcal1:aaaaaaaa}$ in $\eqref{eq:equivcal1:aaaaaaaaa}$, summing over all $B$ blocks and noting that the $B$th block does not contribute, and using the lower bound on \fbox{$\eqref{eq:equivcal1:a}(ii)$}, we get $
 H(W^{[B-1]}|\mathbf Z^{[B]}) \geq (B-1) \times n[R_1 - 2\epsilon]$. \\

\noindent 
\fbox{Case~$2$: $I(X_2;Y_3) + WZ^{Bob} < I(X_2;Z) + WZ^{Eve}$}
\fbox{Case~$2(a)$: $WZ^{Eve}\! <\! WZ^{Bob}$ ($\implies\! I(X_2;Y_3)\! <\! I(X_2;Z)$)}

\begin{figure*}[h] 
\hrule
Equivocation Calculation \fbox{Case~$2$: $I(X_2;Y_3)+WZ^{Bob} < I(X_2;Z)+WZ^{Eve}$ }
 \begin{align}
H(W^{[B-1]}|\mathbf Z^{B}) 
 &=H((\mathbf X_1, \mathbf X_2, \mathbf Z)^{[B]}) + H(W^{[B-1]}|(\mathbf X_1, \mathbf X_2, \mathbf Z)^{[B]}) - H((\mathbf X_1, \mathbf X_2)^{[B]}|W^{[B-1]}, \mathbf Z^{[B]}) - H(\mathbf Z^{[B]}) \notag \\ 
 &\geq H((\mathbf X_1, \mathbf X_2, \mathbf Z)^{[B]}) - H((\mathbf X_1, \mathbf X_2)^{[B]}|W^{[B-1]}, \mathbf Z^{[B]})- H(\mathbf Z^{[B]}) \notag \\
 &=\underbrace{\sum_{j=1}^B \underset{T_j}{H((\mathbf X_1, \mathbf X_2, \mathbf Z)_j|(\mathbf X_1,\mathbf X_2, \mathbf Z)^{[j-1]})}}_{(a)} 
 -\underbrace{H((\mathbf X_1, \!\mathbf X_2)^{[B]}|W^{[B-1]}, \mathbf Z^{[B]})}_{(b)} - \underbrace{\sum_{j=1}^BH(\mathbf Z_j|\mathbf Z^{[j-1]})}_{(c) 
 } \label{eq:equivcal2:a}
\end{align}
Term \fbox{$\eqref{eq:equivcal2:a}(a)$} is a summation. We expand each individual term $T_j$ of \fbox{$\eqref{eq:equivcal2:a}(a)$} by the chain rule:
\begin{align}
 T_j=\! \underset{(i)}{H(\mathbf X_{2,j}|(\mathbf X_1, \mathbf X_2, \mathbf Z)^{[j-1]})}  +  \underset{(ii)}{H(\mathbf X_{1,j}|\mathbf X_{2,j},(\mathbf X_1, \mathbf X_2, \mathbf Z)^{[j-1]})} 
+\underset{(iii)}{H(\mathbf Z_j|\mathbf X_{1,j}, \mathbf X_{2,j}, (\mathbf X_1, \mathbf X_2, \mathbf Z)^{[j-1]})} \label{eq:equivcal2:aa}
\end{align}
%
\begin{align}
\text{Using (see main text) } \phantom{ww}
\text{Term } \fbox{\eqref{eq:equivcal2:aa}(i)} = nR_2, 
\text{ Term } \fbox{\eqref{eq:equivcal2:aa}(ii)} = n[R_1 + \tilde R_1], 
\text{ Term } \fbox{\eqref{eq:equivcal2:aa}(iii)}= H(\mathbf Z_j|\mathbf X_{1,j}, \mathbf X_{2,j}) \label{eq:equivcal2:aaa}
\end{align}
where the last equality in $\eqref{eq:equivcal2:aaa}$ holds for the same reason as in the evaluation of Term $\fbox{\eqref{eq:equivcal1:aaaa}(c)}$, i.e., channel is DMC. 
\begin{align}
\text{Together with } \eqref{eq:equivcal2:aa}, \text{the above gives } T_j =  nR_2 + n[R_1 + \tilde R_1] + H(\mathbf Z_j|\mathbf X_{1,j}, \mathbf X_{2,j}) \label{eq:equivcal2:aaaa}
\end{align}

Using $\eqref{eq:equivcal1:aaaaaaa}$, $\eqref{eq:equivcal2:aaaa}$ 
in $\eqref{eq:equivcal2:a}$, re-arranging slightly, and with appropriate algebra, we obtain the inequality:
\begin{subequations}
\begin{align}
 H(W^{[B-1]}|\mathbf Z^{B}) &\geq \sum_{j=1}^B n[R_2 + R_1 + \tilde R_1] - I(\mathbf X_{1,j}, \mathbf X_{2,j}; \mathbf Z_j) - H((\mathbf X_1, \!\mathbf X_2)^{[B]}|W^{[B-1]}, \mathbf Z^{[B]}) \\
 &\overset{\eqref{eq:equivcal1:mt} }{\geq} \sum_{j=1}^B n[R_2 + R_1 + \tilde R_1 - \min\{R_2, I(X_2;Z)\} - I(X_1;Z|X_2) - \epsilon] - \underset{(b)}{H((\mathbf X_1, \!\mathbf X_2)^{[B]}|W^{[B-1]}, \mathbf Z^{[B]})} \label{eq:equivcal2:aaaaa}
\end{align}
\end{subequations}

\hrule 
\end{figure*}
%
%
\noindent 
Term \fbox{$\eqref{eq:equivcal2:aa}(i)$}$=nR_2$. Note that $\mathbf X_{2,j}$ is a function of $(\mathbf X_{2,j-1},\mathbf {\hat Y}_{2,j-1})$. $\mathbf {\hat Y}_{2,j-1}$ is determined by $(\mathbf X_{2,j-1}, \mathbf Y_{2,j-1})$, and $\mathbf Y_{2,j-1} \leftrightarrow (\mathbf X_{1,j-1}, \mathbf X_{2,j-1}) \leftrightarrow (\mathbf X_1, \mathbf X_2, \mathbf Z, \mathbf Y_2, \mathbf {\hat Y}_2)^{[j-2]}$ form a Markov chain, implying that $\begin{aligned}[t]
\mathbf X_{2,j} \leftrightarrow (\mathbf X_{1,j-1}, \mathbf X_{2,j-1}, \mathbf Z_{j-1}) \leftrightarrow (\mathbf X_1, \mathbf X_2, \mathbf Z)^{[j-2]}
\end{aligned}$ form a Markov chain. Conditioning on $(\mathbf X_1, \mathbf X_2, \mathbf Z)^{[j-1]}$ reduces the possibilities for $\mathbf {\hat y}_2$ per WZ bin from $2^{n[\hat R - R_2]}$ down to $2^{n[\hat R - R_2 - WZ^{Eve}]}$ which -- by our choices -- is exponentially large in all Case~$2$ regimes. Hence every WZ bin is possible in block~$j-1$, and so is every $\mathbf X_{2,j}$.\\
 
\noindent As in Case~$1$, Term \fbox{$\eqref{eq:equivcal2:aa}(ii)$}$=n(R_1 + \tilde R_1)$. 
%
\noindent
Since $R_2 < I(X_2;Z)$, and $\tilde R_1 \approx I(X_1;Z|X_2)$, and since block~$B$ does not contribute, the RHS of inequality $\eqref{eq:equivcal2:aaaaa}$ becomes
$n(B-1)[R_1 \!-\! \epsilon] \!-\! H((\mathbf X_1,\! \mathbf X_2)^{[B]}|W^{[B-1]},\! \mathbf Z^{[B]})$.
Now we need only show that:
$\begin{aligned}[t]
&H((\mathbf X_1, \mathbf X_2)^{[B]}|W^{[B-1]}, \mathbf Z^{[B]}) 
\end{aligned}$
is ``small''. $\because R_2 < I(X_2;Z)$, conditioning on $\mathbf Z_j$ enables Eve to decode $\mathbf X_{2,j}$ correctly whp using $(\mathbf x_{2,j}, \mathbf z_j) \in \mathcal T^{(n)}_\epsilon$. Given $W_j$, the number of possibilities for $\mathbf X_{1,j}$ reduces from  $2^{n[R_1 + \tilde R_1]} \rightarrow 2^{n[\tilde R_1]}$. Knowing both $\mathbf X_{2,j}$ and $\mathbf Z_j$ enables Eve to decode $\mathbf X_{1,j}$ using $(\mathbf x_{1,j}, \mathbf x_{2,j}, \mathbf z_j) \in \mathcal T^{(n)}_\epsilon$ at a rate $I(X_1;Z|X_2)$, which suffices, since  $\tilde R_1 = I(X_1;Z|X_2) - \epsilon$. By 
Fano: 
$H((\mathbf X_1, \mathbf X_2)^{[B]}|W^{[B-1]}, \mathbf Z^{[B]}) \approx (B-1)\times n\epsilon$. So we finally have: $
H(W^{[B-1]}|\mathbf Z^{[B]}) \geq (B-1)nR_1 - (B-1)n\epsilon
$.\\

\noindent
\fbox{Case~$2(b)$: $WZ^{Bob} < WZ^{Eve}$ }
\fbox{Case~$2(b)(i)$: $I(X_2;Y_3) < I(X_2;Z)$}
The calculation is virtually identical to Case~$2(a)$ and is omitted. 
\\
\noindent

\noindent
\fbox{Case~$2(b)(ii): I(X_2;Z) < I(X_2;Y_3)$}\\
\noindent
We choose $R_2 \in (I(X_2; Z), I(X_2;Y_3))$, $\hat R > I(X_2; Z) + WZ^{Eve}$, and $\hat R - R_2 > WZ^{Eve} > WZ^{Bob}$. As in Case~$2(b)(i)$, neither Bob nor Eve can decode $\mathbf {\hat y}_2$. Our choice of $R_2 < I(X_2;Y_3)$ ensures that Bob can decode $\mathbf x_2$. Since both $R_2 > I(X_2;Z) \text{and } \hat R > I(X_2;Z)+WZ^{Eve}$, Eve cannot decode $\mathbf x_2$ \textit{uniquely}. Whether Eve decodes $\mathbf x_2$ nonuniquely or treats it as noise depends on two further sub-cases, depending on whether $I(X_2;Y_3) < \text{ or } > I(X_2;Z|X_1)$. 
Expanding by the chain rule, term \fbox{$\eqref{eq:equivcal2:aaaaa}(b)$}
$\begin{aligned}[t]
 &=\sum_{j=1}^BH((\mathbf X_1, \mathbf X_2)_j|(\mathbf X_1, \mathbf X_2)^{[j-1]}, W^{[B-1]}, \mathbf Z^{[B]})\\ 
 &\leq \sum_{j=1}^BH(\mathbf X_{1,j}, \mathbf X_{2,j}|W_j, \mathbf Z_j)
\end{aligned}$.

$W_j$ in the conditioning reduces the possibilities for $\mathbf X_{1,j}$ from $2^{n[R_1 + \tilde R_1]} \rightarrow 2^{n\tilde R_1}$. There are $2^{nR_2}$ possibilities for $\mathbf X_{2,j}$. Conditioning on $\mathbf Z_j$ reduces the possibilities for the pair $(\mathbf X_{1,j}, \mathbf X_{2,j})$ from $2^{n[\tilde R_1 + R_2]} \rightarrow 2^{n[\tilde R_1 + R_2 - I(X_1,X_2;Z)]}$. 
So we have: $H(\mathbf X_{1,j}, \mathbf X_{2,j}|W_j, \mathbf Z_j) \approx n[\tilde R_1 + R_2 - I(X_1,X_2;Z)]$. Using this in $\eqref{eq:equivcal2:aaaaa}$, and since $R_2 > I(X_2;Z)$ and block $B$ does not contribute, we obtain: 
$\begin{aligned}[t]
 H(W^{[B-1]}|\mathbf Z^{[B]}) \geq (B-1)nR_1 - (B-1)n\epsilon
\end{aligned}$
\section{Conclusion and Future Work}
\label{sec:conc}

\noindent
It remains to be proven 
that our scheme achieves a higher secrecy rate than \cite{LaiElGam08}. NF achieves secrecy improvement by attacking both Bob and Eve. Since Eve is affected more, Bob's secrecy rate can improve in certain regimes. In our scheme, Alice has greater control over the relay's channel codeword in the next block. Our scheme's decoding delay of two blocks is to be preferred to \cite{LaiElGam08}'s backward decoding delay of $B$ blocks. Lastly, our MBEq calculation is, besides \cite{KIyer2018} by the author, one of the first of its kind -- all other MBEq calculations, such as \cite{DaiYuMa16}, to the best of our knowledge, were made in the context of LM-NNC without WZ binning. Thus our equivocation calculations may be of independent interest. The next problem to be tackled is the four node dedicated relay broadcast channel with mutual secrecy requirement where the relay is ``strong'' with respect to one receiver and decode-forwards its intended message, and ``weak'' (or possibly untrusted) wrt the other receiver, and applies a version of compress-forward to its intended message. We foresee substantial technical challenges, but believe the effort will be worth it.

\section{Acknowledgements}
The author is grateful for the support of the Bharti Centre for Communication, IIT Bombay.

%
\nocite{*}
\bibliographystyle{IEEEtran}
\bibliography{./references2}


\end{document}